# THE UNIVERSAL EVOLUTION AND THE ORIGIN OF LIFE

Gennady Shkliarevsky

**Abstract:** The origin of life occupies a very important place in the study of the evolution. Its liminal location between life and non-life poses special challenges to researchers who study this subject. Current approaches in studying the origin and evolution of early life are reductive: they either reduce the domain of non-life to the domain of life or vice versa. This contribution seeks to provide a perspective that would avoid reductionism of any kind. Its goal is to outline a frame that would include both domains and their respective evolutions as its particular cases. The study examines the main theoretical perspectives on the origin and evolution of early life and provides a constructive critique of these perspectives. An objective view requires viewing an object or a phenomenon from all available points of view. The goal of this contribution is not to prove the current perspectives wrong and to deny their achievements. It seeks to provide an angle that would be sufficiently wide and would allow synthesizing current perspectives for a comprehensive and objective interpretation of the origin and evolution of early life. In other words, it seeks to outline a frame for an objective view that will help understand life's place within the universe.

**Key words:** origin of life, prokaryotes, eukaryotes, multicellular life, process of creation, evolution, conservation, and levels of organization.

## Introduction

Life is perhaps the most intimately familiar experience that all humans have. It begins even before we are born. Yet when asked what life is, most of us will be at a loss for an explanation and will be able to offer little more than vague intuitions. For centuries scientists have reflected on life, but a consensus on what life is or how it originated has emerged.[1] Despite its proximity and familiarity, life has been and remains a mystery that defies our imagination and challenges our capacity for rational explanation.

Decades of research have produced an impressive array of ingenious theories and sophisticated experiments; numerous facts about living organisms have become known. Scientists can describe in minute detail all aspects of a living organism, but are unclear about the most important distinction between life and non-life. Yet they have so far failed to provide a definitive solution of the problem of life and its origin. As Nick Lane summarized in the preface to his book *The Vital Question: Energy, Evolution, and the Origins of Complex Life*:

> There is a black hole at the heart of biology. Bluntly put, we do not know why life is the way it is.[2]

A group of researchers led by Eloi Camprubí expresses a similar sentiment: "[I]t is clear," they write, "that we are still far from an answer on how life emerged on Earth, let alone on life's place within the Universe."[3]

The discrepancy between the accumulated knowledge and a lack of the definitive understanding of life is frustrating. Some researchers are even skeptical about the very possibility of having "any scientifically relevant language" that can capture the complex and multifarious phenomenon of life.[4] They claim that we may never know answers to the most fundamental questions related to life. Moreover, they argue that such answers lie beyond our very capacity to know, thus setting limits to human knowledge.

Many, however, find this verdict hard to accept. They believe that the answer to fundamental questions about life require new approaches and new ideas. In his review of David Deamer's book *Assembling Life: How Can Life Begin on Earth and Other Habitable Planets* William Bain opines:

> . . . the Origin of Life field as a whole is unconvincing, generating results in Toy Domains that cannot be scaled to any real world scenario. I suggest that, by analogy with the history of artificial intelligence and solar astronomy, we need much more scale, and fundamentally new ideas, to take the field forward.[5]

The quest for a new approach motivates this essay. It will examine the theoretical perspectives on the origin and evolution of early life and will try to formulate a new and inclusive theoretical perspective that will open new possibilities for addressing the old questions: How did life originate and come to dominate Earth?

**General Theories of the Origin of Life**

*Definitions of Life*

An explanation of the origin of life depends to a significant degree on the way that researchers define life. Current definitions vary widely both in complexity and sophistication. As has already been mentioned, there is no consensus on what life is and what distinguishes it from non-life.

Some definitions are simple, while others are more complicated. Carl Sagan, for example, defined life as "a system capable of evolution by natural selection."[6] NASA's working definition is very similar. Life, it says, "is a self-sustained chemical system capable of undergoing Darwinian evolution."[7] Theoretical biologist Claus Emmeche provides a lengthier (and fancier) definition. He views life in semiotic, or rather biosemiotic terms: "Life is the functional interpretation of signs in self-organizing material 'code-systems' that construct their own 'um-welts.'"[8] Robert Shapiro and Gerald Feinberg think of life in terms of systems theory. They understand life as "a highly ordered system of matter and energy characterized by complex cycles that maintain or gradually increase the order of the system through an exchange of energy with its environment."[9]

Other examples include Francisco Varela and Humberto Maturana, famous Chilean biologists, who see life through the prism of autopoiesis—the term that they invented and that focuses on the capacity of systems to reproduce, maintain, and renew themselves. A living organism is, in accordance with their perspective, "the minimal living organization" that is defined as "a network of processes of production (synthesis and destruction) of components such that these components: (i) continuously regenerate and realize the network that produces them, and (ii) constitute the system as a distinguishable unit in the domain in which they exist."[10] A group of researchers led by Ruiz-Murazo, characterizes a living being as "any autonomous system with open-ended evolutionary capacities."[11] In yet another definition, life is "a self-sustained, dynamic, non-equilibrium, extended replicative chemical network, which was initiated by the chance emergence of a simple, but persistent, replicative chemical system."[12]

There seems to be as many definitions of life as there are researchers that work in this field. Despite their diversity, these definitions share some points. They all consider autonomy to be a principal feature of life. Another characteristic feature that many researchers emphasize is life's capacity for "open-ended evolution." Most of them view the evolution through the prism of Darwinian theory that explains the evolution in terms of variations and natural selection.

It is worth pointing out that Darwin developed his theory primarily with biology in mind and this theory has been primarily applicable to the biological realm. Yet the origin of life involves both the inanimate as well as the animate world. For this reason, the Darwinian approach toward the evolution is perhaps the most problematic aspect of many current definitions of life

*Current OOL Perspectives*

1. The "privileged function" perspective

One popular perspective on the origin of life (OOL) is the so-called "privileged function" perspective. Kathryn Lanier and Loren Williams define "privileged function" as "an extant biological function that is excised from its biological context, elevated in importance over other functions, and transported back in time to a primitive chemical or geological environment."[13] In other words, this perspective tries to locate what belongs to the more powerful level of organization—i.e., life—in the inorganic realm that represents a less powerful level of organization. As Lanier and Williams further explain, privileged functions "are considered by OOL model builders to be so essential and fundamental to life that they must also be a requisite for the origin of life."[14]

The majority of the "privileged function" models fall into one of the two major groups: the "gene first" approach and the metabolism first approach. The proponents of the "gene-first" approach focus on autocatalytic self-sustaining chemical reactions that they see as fundamental to all living organisms. They argue that life has its roots in the capacity of molecules to create functionally closed and self-sustaining system characterized by internal autocatalytic operations and functions. Such systems, in their view, eventually gave rise to the intricate RNA-DNA-protein molecular machinery that made life possible. From this proposition they draw a conclusion that if autocatalysis sustains life, then life itself is little more than an autocatalytic set and its spontaneous emergence from pure chemistry is less improbable than one might think.[15]

The "gene first" hypothesis maintains that life began with the emergence of RNA molecules. It sees the capacity of RNA for template-based self-replication and the ability to catalyze other reactions as playing the decisive role in the emergence of life; hence the name for this hypothesis is RNA Worlds.[16] Clay Worlds is a subgroup of RNA Worlds that focuses on a variety of mineral surfaces as sites for synthesizing bio-molecules capable of replication.

2. The "metabolism first" perspective

"Metabolism first" is also a popular approach in the "privileged function" category. This model emphasizes metabolism as the principal function of living organisms. The "metabolism first" approach offers a variety of models. One popular model is the so-called "managed metabolism" hypothesis. This hypothesis argues that in order to make the transition from non-life to life, self-producing systems of chemical molecules must overcome the so-called "cooperation barrier" that prevents the cooperation in their unmanaged, self-organizing, autocatalytic networks. These autocatalytic networks of molecules have a variety of the co-called "free riding" processes that work against cooperation and they need to be managed. The proponents of this hypothesis identify the emergence of the system that manages the un-regulated processes and turns them into a system of orderly metabolism with the onset of life.[17] The mechanism that provides this regulation and overcomes the "cooperation barrier" is the DNA/RNA apparatus since effective regulation requires information and memory that stores this information.[18] An interesting feature of this model is that it sees the primary role of DNA/RNA in regulation, rather than in replication.

The Thermal Vent Worlds hypothesis is a subset of the "metabolism first" category. It focuses on the process of harvesting energy form chemical gradients that takes place at submarine hydrothermal vents. These vents became the site of abiotic synthesis of simple organic molecules and later polymers.[19]

3. The Membrane Worlds

The Membrane Worlds is yet another popular hypothesis in the "privileged function" category. The "privileged function" in this model is compartmentalization. Its proponents focus on the role of membranes in the emergence of life. In their view, the formation of compartments was highly consequential as it produced cascading structures from simple to complex systems.[20] In several "privileged function" models, compartmentalization is a secondary privileged function, used as an adjunct to a primary privileged function.[21]

### 4. The continuity perspectives

There are some hypotheses that emphasize continuity, rather than discontinuity, in the emergence of life.  Such hypotheses base their explanations on general universal laws that operate in the physical world.  In other words, they try to explain biology from physics.[22]  As d'Arcy Thompson—one of the early representatives of this trend—has argued, biologists tend to overemphasize the role of the biological evolution over that of physical laws in the origin and growth of life.[23]  Nigel Goldenfeld and Carl Woese, two relatively recent contributors to this line of thinking, argue:

> Evolution is the fundamental physical process that gives rise to biological phenomena . . . We discuss how condensed matter physics concepts might provide a useful perspective in evolutionary biology, the conceptual failings of the modern evolutionary synthesis, the open-ended growth of complexity, and the quintessentially self-referential nature of evolutionary dynamics . . . Indeed, the very existence of biological phenomena is an expression of physical laws that represent a new asymptotic realm in non-equilibrium statistical physics . . . *Life is physics and evolution is a collective phenomenon far from equilibrium*.[24]

Self-organization is another common theme among explanations that emphasize continuity.[25]  Its proponents see the emergence of life to be primarily a result of intrinsic general properties of matter, such as self-organization, rather than a product of natural selection.  Concepts related to self-organization initially found their application in chemistry and physics, not biology. Their application to various morphogenetic problems in biology is a later development.  The use of this approach to explain the origin of living cells did not begin until the last century.[26]

According to the proponents of this perspective, the capacity for self-organization played a decisive role in the origin and evolution of early life forms.[27]  Many of them share the view of Roland Wedlich-Söldner and Timo Betz who argue that self-organization "lies at the heart of the robustness and adaptability found in cellular and organismal organization, and hence constitutes a fundamental basis for natural selection and evolution."[28]

Another trend in the continuity category emphasizes the connection between entropy and the origin of life.  Entropy production is central to this trend.[29]  The fact that every human generates 6,000 times more heat per kilogram of mass than the Sun leads, for example, Roy Murphy to conclude that "the conversion of pure energy into more disordered energy such as thermal, kinetic and chemical is the whole reason for our being."[30]

Jeremy England, a physicist from MIT and one of the recent contributors to this line of thinking,  According to England,

> The formula, based on established physics, indicates that when a group of atoms is driven by an external source of energy (like the sun or chemical fuel) and surrounded by a heat bath (like the ocean or atmosphere), it will often gradually restructure itself in order to dissipate increasingly more energy.[31]

England argues that the origin and subsequent evolution of life follow from the fundamental laws of nature and "should be as unsurprising as rocks rolling downhill."

*Critique of the Current OOL Perspectives*

1. The flawed strategy

Many current OOL perspectives use the same common strategy. They focus on one particular property (function or feature) in living organisms that they posit as fundamental to life. They view the rise of this property in the inanimate world as the beginning of life and try to reconstruct the process of its emergence.

The logic of this strategy is essentially circular and tautological. Rather than proceed from causes to effects, this strategy takes the results and tries to find their causes. No wonder that attempts to locate a property that belongs to a more powerful level of organization in the level that is less powerful end up in failure.[32] A weaker level of organization simply cannot sustain a feature that requires a more powerful level of organization. As Addy Pross, among others, has concluded: "The origin of life (OOL) question might be considered physical organic chemistry's ultimate challenge, yet despite continuing efforts over close to a century, the problem remains unresolved."[33]

2. The problem of subjectivism and arbitrariness

There is another problem with the strategy of proceeding from results to causes, rather than vice versa. This strategy involves constructing theories on the basis of some assumption that is regarded as self-evidently true—for example, the assumption that metabolism or replication is the fundamental function of the organism—and then using this assumption as the organizing principle for the theory.

Assumption is essentially an educated hunch; and as all hunches go, assumption, educated or not, is also to a significant degree subjective and arbitrary. The way scientific methodology addresses this problem of subjectivity and arbitrariness is by subjecting the foundational organizing principle of a theory to the test of rational justification and empirical verification. This test is not an ironclad proof, but it does show that the principle on which a theory is built is not entirely arbitrary. The application of rules of logic, reason, and empirical proof shared by other researchers shows that the author at least to some extent controls the theory, rather than the theory controls the author.[34]

None of the perspectives on the origin of life offer rational justification or empirical verification to support the assumptions used as their foundational organizing principle. For example, metabolism and replication are very important biological functions, but there is absolutely no reason to assign a "privileged" status to either of them. Despite their theoretical sophistication and an impressive cache of empirical data, models that do not provide rational justification or empirical verification for their foundational principles cannot be considered objective and they cannot effectively respond to criticism for being arbitrary and subjective.

### 3. The randomness problem

Most current OOL perspectives invoke chance or coincidence in their explanations of the origin of life. Sean Carroll, a well-known evolutionary biologist, refers to the emergence of life as "the mother of all accidents" and "the accident of all mothers."[35] The problem with invoking chance or coincidence is that one cannot really demonstrate their existence (in other words, provide a rational justification for its existence). Genuine randomness may or may not exist in nature, but there is really no way to prove either its existence or, for that matter its non-existence. For example, a set of numbers may show randomness, but there is no guarantee and no proof that if we extend this set indefinitely, it will not reveal some underlying order or pattern.[36]

There are a growing number of biologists who dispute the critical assumption of randomness. Kevin Laland et al., for example, find that "much variation is not random because developmental processes generate certain forms more readily than others"[37] Christian de Duve the author of *A Guided Tour of a Living Cell*,[38] also offers an objection to using chance in an explanation. In one of his widely publicized quotes, he writes:

> If you equate the probability of the birth of a bacteria cell to chance assembly of its atoms, eternity will not suffice to produce one . . . Faced with the enormous sum of lucky draws behind the success of the evolutionary game, one may legitimately wonder to what extent this success is actually written into the fabric of the universe.[39]

Many OOL perspectives rely on Darwinian theory. This theory does offer a proof that mutations exist. However, it does not and cannot demonstrate that these mutations are random. The lack of proof means in this case that there is at least a fifty per cent chance that initial variations may be a result of some underlying order, which would be an anathema to the Darwinian model. Again, this is not to say that initial variations are not random. This is to say, however, that there is no way of proving their randomness and, therefore, it must be put in doubt.

The lack of rational justification and empirical verification for the existence of chance indicates that a theory based on the assumption of the existence of chance is not under control of reason and, therefore, is subjective and arbitrary. A possibility of such theory controlling us, rather than we controlling it, is dangerously strong. Chance and coincidence are not really conducive to rational explanation as they set a limit to our knowledge. In addition, the distance that separates researchers from the time when life emerged offers no opportunity for empirical verification either. For example, we have no empirical proof for the existence of hypothetical common ancestors, yet many scientists use these entities to bolster their theories.

### 4. The continuity/discontinuity problem

The use of chance and coincidence in explanations of the origin of life also points to a deeper problem: the failure by the current perspectives to resolve the problem of continuity and discontinuity. An overwhelming majority of biologists accept the evolution as a fact. They also accept that life has its roots in inorganic matter. In other words, they de facto recognize that the origin of life involves continuity. However, they also recognize that life represents a totally new level of organization that did not exist prior to its emergence. Therefore, they also recognize that the origin of life involves discontinuity. However, they do not explain how the two are interrelated. Moreover, they do not even raise this issue, accepting the existence of both continuity and discontinuity as a given. The solution of the fundamental problem of the origin of life is not possible without explaining how continuity and discontinuity are related. By invoking chance the current perspectives on the origin of life do not even have to explain this relationship. They simply evade this subject altogether.

### 5. The selection problem

As has already been mentioned, many current OOL models use "privileged function" as a category fundamental to their interpretation. In addition, they commonly use "universal common ancestor" as another category that they deem to be central to their explanations.[40] They define "universal common ancestor," or "the last universal common ancestor" (LUCA), as a population of the most recent organisms to which they trace the origin of all species. The use of both categories creates confusion as to what was the unit of selection in the origin of life: function or form (common ancestor).

The recognition that life emerged from the inorganic world means that the origin of life was a result of the evolution. The Darwinian model to which many OOL perspectives subscribe, works on selection. If selection is involved, there must be also a unit of selection that the evolution conserves. The identification of the unit of conservation is critically important in explaining the evolution. The fact that that the current OOL perspectives use two very different units of selection—function and organism—creates confusion and makes the reconstruction of the evolution difficult.

Function and organism are very different from each other. Function is broader than organism. Many organisms that are very different from each other may have similar functions. The fact that function relates to many different organisms works better with evolutionary processes that operate on a broad scale that embraces multiple organisms. Conservation of a particular function will have a much broader scope that will cover many organisms.

Function is a form of action. It is dynamic in its nature. So is the evolution that is also characterized by action and dynamism. If dynamism is the nature of the evolution, then the evolution is primarily about conserving its own dynamism. As it evolves, the evolution becomes increasingly more dynamic. Since dynamism is so central to both the function and the evolution, conserving function is more compatible with the nature of the evolution than conserving an organism, or form, that is static in its nature.

The history of sexual reproduction may help to illustrate this point. Sexual reproduction is not the oldest form of reproduction. It goes back only to the emergence of eukaryotes.[41] Yet today, sexual reproduction is ubiquitous; it is the dominant form of reproduction in nature.

The dominance of sexual reproduction in nature puzzles researchers.[42] They still cannot explain why the evolution preferred sexual reproduction to asexual one. Some argue that sexual reproduction has important advantages over asexual one. They point, for example, to the capacity of sexual reproduction to produce beneficial mutations and improve genotypes, which advances the evolution. Their critics object by arguing that sexual reproduction can also produce harmful mutations and disrupt adaptive genotypes during independent assortment and crossing over of genes.[43] Also, sexual reproduction involves a huge investment in terms of resources. After all, bacteria do not have to search for a mate. They simply divide in two. No need to invest energy in finding a partner, fertilizing an egg, and joining two genomes.[44]

The prominent British evolutionist Richard Dawkins compares sexual reproduction to a game of roulette "in which the player throws away half of his chips at each spin." "The existence of sexual reproduction really is a huge paradox," Dawkins concludes.[45] The evolutionary biologist Graham Bell is even more direct in his assessment:

> Sex … does not merely reduce fitness, but halves it. If a reduction in fitness of a fraction of one percent can cripple a genotype, what will be the consequence of a reduction of 50 per cent? There can be only one answer: sex will be powerfully selected against and rapidly eliminated wherever it appears. And yet this has not happened.[46]

One may agree or disagree with either side in this controversy. However, there is clearly no definitive answer as to why the evolution favored sexual reproduction. Given what we know, it remains unclear why the evolution would not treat both forms of reproduction equally. As Lauren Hurst asks, why not have the best of both worlds and benefit equally from what both have to offer?[47]

If we think about both forms of reproduction, no clear advantages emerge from either sexual or asexual reproduction. Both have advantages and disadvantages. However, if one thinks about the reproductive function itself, rather than its specific form, the evolution shows preference for sexual reproduction as the more effective way of conserving the reproductive function itself. Sexual reproduction conserves the function of reproduction better than the asexual form of reproduction.

Function is about action. In order to conserve action, it has to be enacted. The more often a function is enacted, the better it is conserved. Sexual reproduction offers more opportunities to stimulate and activate the reproductive function. In a population of sexually differentiated organisms the number of real and potential sources of stimulation of the reproductive function increases exponentially as individual organisms interact with each other. In other words, sexual reproduction increases the stimulation and activation of the reproductive function and, thus, conserves it better than asexual reproduction does since asexual reproduction does not offer multiple external sources of reproductive stimulation.

The above is a simple illustration why function, rather than form, is a more likely object of conservation in the evolution. However, one should keep in mind that since life originated in the inorganic world, the object of conservation in this case should not be a biological function. It can only be a function that extends to both worlds.

   6. The failure to produce a synthesis

Objectivity requires viewing an object or a phenomenon from all sides. The current perspectives on the origin of life do not observe this condition. They all have their specific focus, use their specific facts, and interpret these facts in their own way. Each perspective uses its own specific prism. Each has its own specific angle from which it views the origin of life. As a result, no matter how valid the points that each perspective brings up, each is, to one degree or another, incomplete and, therefore, subjective. In order to solve the problem of the origin of life, the field needs a comprehensive approach and a frame that would be broad enough to include all perspectives as its particular cases—cases that are valid under specific conditions or assumptions.

**Deliberating the Main Stages in the Origin and Evolution of Early Life**

The universally accepted view in the scientific community is that the origin and evolution of early life is a process that includes several stages. The creation of organic molecules was the first important stage in this process. It opened the path for the subsequent stages: prokaryotic cells (bacteria and archaea), modern eukaryotic unicellular organisms, and finally multicellular life. Each of these stages represented a new and increasingly more powerful level of organization. Each has emerged from the level of organization that preceded it; and each has given rise to the next new and more powerful level of organization.

*The Origin of Organic Molecules*

The origin of organic molecules is still a much-debated issue that so far has no definitive solution. There are numerous hypotheses as to how organic molecules may have emerged. A detailed examination and critique of these hypotheses is beyond the scope of this study and a general discussion will have to suffice.

Most hypotheses regarding the origin of organic matter focus on autocatalytic chemical reactions. The proponents of this approach see a set of autocatalytic and self-sustaining chemical reactions as fundamental to all living organisms. The conclusion that they infer from this observation is that life has its roots in the capacity of matter to create functionally closed and self-sustaining system characterized by internal autocatalytic operations and functions. In their view, such systems gave rise to the intricate RNA-DNA-protein molecular machinery that is fundamental to replication and life.[48] This proposition leads them to conclude that if autocatalysis sustains life, then life itself is little more than an autocatalytic set and its spontaneous emergence from pure chemistry is less improbable than one might think.[49]

This general perspective invokes chance. The common argument for most hypotheses in this category is that the right combination of necessary ingredients and conditions of early Earth made the emergence of organic molecules possible. The classic Oparin-Haldane hypothesis, more commonly known as "the primordial soup" hypothesis, has been dominant over the last century and still retains its popularity.[50] The Miller-Urey experiment on the formation of nucleobases is perhaps the best-known experimental attempt to prove this hypothesis.[51] This perspective contends that any theory of abiogenesis must take into account the condition of the Earth about 4.5 billion years ago. They argue that organic molecules were a product of something similar to the Miller-Urey process that took place under the conditions of cooled Earth with its abundance of water. According to this scenario, Earth's oceans became a huge pool teeming with a variety of molecules. Given enough time, they combined by pure chance to form first organic entities that were capable of replication.[52] This original hypothesis has subsequently evolved into what is known today as the RNA hypothesis that is still very common among biologists and is used in many textbooks.[53] The big attraction of this hypothesis is its similarity with Darwin's "warm little pond" proposal. [54]

Over the years researchers have added more models describing the emergence of the organic world.[55] Some of them point to various mineral surfaces—such as, for example, clay--as a possible catalyst for the origin of organic molecules and life.[56] Others propose deep-see thermal vents as possible sites of the origin of organic matter.[57]

These hypotheses on the origin of organic molecules have several major and unresolved problems. For one thing, their main focus is not so much on causes as on factors that facilitated the emergence of the organic world. Also, these hypotheses emphasize discontinuity in this process and play down continuity. There is no justification for prioritizing discontinuity over continuity. Just like discontinuity, continuity is of paramount importance. Without continuity, there would be no evolution.

As a result of their emphasis on discontinuity, these hypotheses have to rely on chance and coincidence as explanatory factors in the emergence of life. The big weakness of such hypotheses is that their assumption about the role of chance and coincidence cannot pass the test of rational justification and empirical verification, which makes them vulnerable to criticism for being subjective and arbitrary.

Analyses of the chemical composition of meteorites and comets have recently revealed traces of organic matter,[58] which led to the formulation of a number of hypotheses that suggest that organic molecules may have originated somewhere else in the universe and later came to Earth in the shower of meteorites that pelted the planet.[59] These hypotheses of interplanetary origin of organic molecules also do not solve the problem of origin; they merely move this problem to some extraterrestrial location. By arguing for an extraterrestrial origin, they open possibilities that the origin of life does not really need the special conditions that existed on Earth. They suggest that life could have emerged under conditions that are very different from those on our planet, which opens a possibility that factors other than terrestrial ones may have been involved. Reuben Hudson and his colleagues, among others, argue that there is "no need for special conditions of Earth to produce life" and "life can emerge elsewhere in the cosmos."[60]

*The Emergence of Prokaryotic Cells*

The currently prevailing theories about the emergence of prokaryotic cells—bacteria and archaea—argue that these first types of cells emerged from a common ancestor—a protocell, or protobiont. A protobiont is an aggregate of abiotically produced organic molecules capable of replication surrounded by a membrane or a membrane-like structure and having a capacity for rudimentary metabolism. Membranes performed a very important role. They separated the internal space from the external environment and made possible to keep close by the products of the genetic material. Membranes evidently offered a huge advantage as the enclosed replicators quickly outcompeted "naked" replicators.[61]

Biologists who study the emergence of protobionts are still unclear as to how this structure came about. They recognize that they know "disappointingly little about how any of these remarkable biological innovations came about." As Greg Fournier, a geobiologist at the Massachusetts Institute of Technology and a leading authority in the field, acknowledges: "It's very hard to infer even the relative ordering of evolutionary events before the last common ancestor." Eric Gaucher, a biologist at the Georgia Institute of Technology in Atlanta, agrees that there is "a huge chasm between the origins of life and the last common ancestor."[62]

Some researchers believe that three main features—genetic reproduction, metabolism, and membranes—emerged independently and later combined to form a permanent structure. Others contend that one of them—for example, genetic replication—emerged earlier and helped generate the other two features.[63] Be that as it may, most researchers recognize that there is a non-trivial chicken-and-egg problem associated with the origin of the common ancestral cell.[64] The fact is that in order to do its work, the genetic machinery needs a membrane. Membranes are loaded with proteins. In fact, proteins account for roughly half the mass of most cellular membranes.[65] Thus, the genetic machinery cannot produce proteins without a membrane, but the synthesizing of membranes requires proteins that can only be produced by the genetic machinery of the cell.[66]

There are also additional problems that plague the dominant scenario on the origin of prokaryotic cells. Most, if not all of them, invoke chance or coincidence to explain the origin of major cell components. Not only that, but they also invoke chance or coincidence to explain why these components came together to form a cell. As has been explained earlier, invoking chance even once is highly problematic; invoking it twice to explain the same phenomenon makes an explanation very questionable.

Not all biologists subscribe to the scenario that the three major components of life formed separately. In fact, more and more scientists, including David Deamer and Jack Szostack, adhere to the so-called "everything-first" hypothesis. John Sutherland at the MRC Laboratory of Molecular Biology in Cambridge, UK, and his team have concluded that all the cellular subsystems "could have arisen simultaneously through common chemistry."[67]

Atsushi Kamimura and her colleague Kunihiko Kaneko, both from the University of Tokyo, have proposed what they call the cluster model.[68] They argue that in the case of the origin of cells, the evolution acted on clusters of interacting organic molecules that gradually increased their complexity and improved copy fidelity. In other words, prokaryotes and all their important features emerged as a product of clusters of interacting organic molecules. The formation of such clusters played a singularly important role in the origin of prokaryotes. This approach helps solve several problems at the same time: the emergence of heredity, metabolism, and compartmentalization.[69]

The path of the "everything-first" hypothesis is still a work-in-progress. Certainly more theoretical and empirical work is needed to flesh out this perspective, but it does show promise. The growing number of fossil records reveals aggregations of organic material. These multi-layer mats varied in thickness and were held together by a sticky glue-like substance. Such mats could serve as sites for the emergence of the level of organization that could simultaneously give rise to structures with combined differentiated components: genetic replication, membranes, and metabolism.[70] The advantage of the "everything-first" hypothesis is that instead of three largely independent processes, this hypothesis envisions one process that produces all major cellular components at the same time and there is no need to invoke chance to explain why they combined together to form a cell.

*From Prokaryotes to Eukaryotes*

The next important stage in the evolution of early life after the rise of prokaryotic cells was the formation of modern eukaryotic cells. Most researchers recognize that this stage of the evolution dramatically departs from the Darwinian model. According to the Darwinian model, the driving force of the evolution is competition. By practically universal recognition the rise of eukaryotic cells was due primarily to cooperation. At some point, an archaeon and a bacterium fused together and the two organisms began to live as one.  In this new organism, bacterium became an internal energy producing organ—mitochondrion.  Exactly how this happened is still under debate.  Perhaps the recently discovered earliest closest relative to eukaryotes—*Asgard archaea*—will provide at least some necessary answers.[71]

There are basically two theoretical perspectives on the fusion of archaea and bacteria.  The famous microbiologist Lynn Margolis laid the foundation for one of them in the 1960s.  This theory of endosymbiosis, as Margolis called it, contends that the origin of the modern eukaryotic cell was a result of cell-cell fusion that involved bacteria and archaea.  A cluster of bacteria enclosed in the ancient archaean host transformed into mitochondria—an essential energy producing component of the modern eukaryotic cell.[72]  The adaptation of both components to the new and more powerful structure that they created produced the modern eukaryotic cell, thus opening the path to further evolution**.** [73]

Critics of endosymbiosis have built on the work of those (like Carl Woese, for example) who argued that populations, rather than individual organisms, are a proper unit of conservation in this evolution.  Woese suggested that biology should conceive the last universal common ancestor, or LUCA, as a multilineage consortium, rather than some ancestral organism.[74]

A group of biologists led Maureen O'Malley and Michelle Leger has formulated a theoretical perspective that differs significantly from endosymbiosis.  In an essay entitled "Concepts of the Last Eukaryotic Common Ancestor" that has appeared in *Nature Ecology & Evolution* O'Malley and her colleagues argue in support of the view that the last eukaryotic common ancestor (LECA) was not a single cell but really a population of genetically diverse cells, none of which had all the characteristics associated with the modern eukaryotes.[75]  "When we're talking about LECA," O'Malley explained in an interview, "we're probably talking about an ancestral state, a genomic state that we don't know was one single cell."[76]  George Mikhailovsky and Richard Gordon also argue that interactions and horizontal gene transfers between archaea and bacteria may explain why the transition from prokaryotes to eukaryotes took over 2.5 billion years.[77]  In other words, this line of thinking suggests that aggregations of prokaryotes had preceded and were an important condition for the emergence of eukaryotes.

The verdict is still out as to which of the two hypotheses offers a better understanding of the process that led to the rise of eukaryotes.  However, while the outcome of this contestation is important, in one sense the outcome makes no difference:  in both scenarios the emergence of eukaryotes was due to cooperation and symbiosis between bacteria and archaea, not competition.  Both perspectives see the emergence of a new and more powerful level of organization that made the archaeon-bacterium combination possible.  Both hypotheses—whether the rise of eukaryotes involved symbiosis of individual cells or whether the origin occurred in aggregations of cells—talk about the emergence of a new level of organization.  Their only disagreement is about how this process occurred.  As Ursula Goodenough and Joseph Heitman summarize:

The transition from prokaryote to protoeukaryote to the last eukaryotic common ancestor (LECA) entailed conservation, modification, and reconfiguration of preexisting genetic circuits via mutation, horizontal gene transfer (HGT), endosymbiosis, and selection.[78]

*Multicellular Life*

For over three billion years since life emerged on our planet, unicellular organisms dominated the terrestrial environment. Then about 600 to 800 million years ago life made another major evolutionary leap when cells formed mutlicellular organisms. The magnitude and complexity of this transition are hard to overestimate. In order to create this new level of organization, cells had to learn to cooperate. They had to create common structures in which they divided labor and differentiated their functions. They produced new and more complex ways of communicating with each other and sharing resources. They learned to protect the common structure of which they were integral parts and procure means necessary for its survival. This epic transformation made life on Earth dramatically different and set a new course for the evolution that created the world in which we live today.[79]

The beginning of multicellularity was rather modest. Then about 540 million years ago, multicellular life erupted into what is known as the Cambrian explosion when a massive number of animal prototypes emerged. Practically every animal that has ever lived since that time until today is a variation on one of the body plans that emerged during the Cambrian explosion.

New details about the transition are still coming out and many are still unknown. However, some fundamental conclusions have already emerged. Regardless of the specifics for different lineages and clades, the path of emergence of all multicellular organisms involved a similar series of steps from the unicellular mode of existence to the colonial or siphonous mode and finally to the parenchymatous one. As Karl Niklas, a prominent contributor to the field, concludes:

> How exactly steps such as cell-to-cell adhesion or communication were achieved in plants, animals, fungi, and algae differs among the major eukaryotic clades, yet an important aspect is that these multicellular organisms all went through a similar series of steps on their way to becoming multicellular, functional organisms.[80]

In all instances cell-to-cell adherence and communication were the *conditio sine qua non* that formed the foundation of multicellularity across all clades.[81]

Many what and why questions about the emergence of multicellular life are still a subject of heated debates.[82] Some researchers argue that the emergence of aggregations (colonies) of unicellular organisms led to their specialization and differentiation of functions.[83] Others see it as a product of the division of individual cells.[84]

In contrast to the theories of the emergence of life, theories of the emergence of multi-cellular organisms do not invoke chance or coincidence.[85]  With much evidential support they all argue that it was inevitable.  Karl Niklas and many other biologists show that the transition to multicellularity occurred on at least 20 different occasions in different lineages from algae to plants to fungi.[86]

Researchers still debate why unicellular organisms formed colonies.  Why single cells began to live together when this common living clearly put them at a fitness disadvantage?  Many biologists find the emergence of colonies puzzling from the point of view of Darwinian evolution since such collective living increased competition among cells for resources.  As Pierre Durand, a biologist at Wits University, has pointed out, the solution of this puzzle remains elusive.[87]

A number of researchers point to serious difficulties on the path to multicellularity.  László Nagy, an evolutionary biologist at the Biological Research Centre of the Hungarian Academy of Sciences, observes that for all the benefits of multicellularity its evolution faced major genetic hurdles.[88]  Yet multicellular life did emerge.  Moreover, the evolutionary histories of some groups of organisms record repeated transitions from single-celled to multicellular forms, suggesting that the hurdles could not have been so high.[89]  Nicole King notes the permeability of the boundary that separates unicellular and multicellular organisms since some unicellular organisms can easily live either as single cells or as multi-cellular colonies.[90]  The fact that multicellular life emerged, endured, and prospered indicates that conservation had a lot to do with its origin.

*Concluding observations*

The discussion of the perspectives on the emergence and evolution of early life warrants some preliminary observations:

1.  The accumulated theoretical and empirical knowledge about the origin and evolution of early life does not fit well with the Darwinian model of the evolution. In explaining the evolution, Darwinists emphasize variability and natural selection.  Yet variability does not appear to be a prominent factor in the origin of life or in the course of life's early evolution.  This fact has led some biologist, and not only biologists, to doubt the role of natural selection during the early stages of life.  John Tyler Bonner, a highly respected figure in the study of early life, is one of the detractors.  Bonner finds intriguing that "extinctions, as in dinosaurs, are common among mega-organisms, and living fossils, such as the horseshoe crab, are rare," which is "in sharp contrast to microorganisms, among which we find living fossils to be common."  Bonner's research led to the conclusion that has raised many brows.  He writes in one of his works:

In unicellular microorganisms, which at one time in early

Earth history were probably the only living eukaryotic forms, *natural selection plays a relatively minor role*, but with size increase, first made possible by the invention of multicellularity, selection plays an increasingly central role in evolutionary change.[91]

A growing number of researchers find that cooperation and symbiosis were much more characteristic for early life than competition. Margaret McFall-Ngai and her collaborators point out that "features once considered exceptional, such as symbiosis, are now recognized as likely the rule, and novel models for research are emerging across biology."[92] Leo Mathieu comes to a similar conclusion with regard to the evolution of prokaryotes. In a chapter characteristically entitled "Original, Non-Darwinian Evolution of Prokaryotes" he writes:

> Associations and mixed communities of prokaryotic strains are the rule. So, on the basis of what is known about genetic exchange, metabolite and enzyme sharing, indications exist that for the overwhelming majority of prokaryotes the selective pressures have favored collaboration, not reciprocal or unilateral destruction. Prokaryotic evolution has been and still is non-Darwinian.[93]

2. The Darwinian theory has originated and found its primary application in biology. There are considerable difficulties with applying Darwinian theory to other fields where evolutionary processes can be observed.[94] Its application to the non-organic domain from which life emerged is particularly problematic. Yet paradoxically, most current approaches continue to reflect Darwinian thinking. For example, many of them maintain that the origin of life involves conservation of a particular function. Yet they derive this function from the biological realm. As a result, they face a formidable task of using a biological function to explain the evolution that begins in the non-organic domain to which this biological function is not really relevant. Indeed, the origin of life may involve the conservation of a particular function, but this function should not be specific to biology; it should be relevant to the evolution of both life and non-life. In other words, it should have a much broader application.

3. As has been repeatedly pointed out, the liberal use of chance and coincidence as an explanatory mode is also a source of concern in theories about the origin and evolution of early life. The invoking of chance or coincidence cannot pass the test of rational justification or empirical verification. Using them every time that discontinuity has to be explained sends a warning about theoretical limitations that make such explanations vulnerable to criticism for being subjective and arbitrary. There is ample evidence showing that continuity and discontinuity play equally important and complementary roles in the evolution.  For this reason, the study of the evolution needs a perspective that would reconcile continuity and discontinuity and would not see them as opposed to each other.  Just as the concept of evolution compels us to view life and non-life as two distinct but closely related stages in the origin of life, it also compels us to view continuity and discontinuity as two equally important and complementary aspects of the same process.

4.  There is also another important concern that relates to all perspectives on the origin and evolution of early life.  They all make valid points that reflect different aspect of this complex process.  An objective view of this process should include all these perspectives as its particular cases.  In other words, a synthesis of all perspectives will really be beneficial for understanding the origin and evolution of early life.  Yet so far the prospects for such synthesis look dim.  As a result, there is a growing awareness of the need for a new angle—one that would bring all perspectives together and open a possibility for a new and comprehensive vision.  One can wholeheartedly agree with William Bains who forcefully writes in his critique of Deamer's book *Assembling Life*:

> I argue that the Origin of Life field as a whole is unconvincing, generating results in Toy Domains that cannot be scaled to any real world scenario.  I suggest that, by analogy with the history of artificial intelligence and solar astronomy, we need much more scale, and fundamentally new ideas, to take the field forward.[95]

**The New Perspective on the Origin and Evolution of Early Life**

The study of the origin and evolution of early life is still an ongoing project.  Researchers have formulated many theoretical perspectives and amassed a great deal of information.  Yet a definitive theory remains elusive.  Many lacunae are yet to be filled and more facts are necessary

It is beyond the scope of the present work to complete this task.  Its goal is much more modest.  It seeks to provide an angle that could serve as a foundation for a frame that would be broad enough to include all current perspectives as particular cases—that is, cases that are true under specific conditions or assumptions.

Although much about the origin and evolution of early life remains unknown, it is not a closed book either.  There is actually quite a lot that we already know and agree.  For one thing, all current perspectives recognize that life represents a new level of organization that did not exist prior to its emergence.  The conclusion that follows from this broad agreement is that the process that resulted in the emergence and evolution of early life involved creation.

The second point that that all researchers agree on is that life has emerged from inorganic matter.  The idea of the inorganic origin of life is widespread among biologists, paleontologists, and researchers from other disciplines.[96]  Darwin's thinking also reflects this intuition.  He was convinced that "the intimate relation of Life with laws of chemical combination, and the universality of latter render spontaneous generation not improbable."[97]  If the process that led to the origin and evolution of early life has its roots in the inorganic domain, then this process is the bridge that connects the two domains.  The processes that create radical novelty either in the domain of life or in the domain of non-life must have important common features, which suggests a possibility for a common perspective on the process of creation that operates in both domains.

*The Process of Creation and the Universal Evolution*

As has already been pointed out, life represents a radical novelty—that is, it represents a level of organization that did not exist prior to its emergence.  We associate radical novelty with creation.  Therefore, we have every reason to view the emergence of life as a result of the process of creation.

The creative act that led to the emergence of life is not unique to biology, or in fact any other field.  There are many objects and phenomena in our universe that represent radical novelty—that is, they did not exist prior to their emergence.  The process of creation is as common in the non-living world as it is in life world.  The results of this process are all around:  from particles and atoms, to molecules and clouds of molecules, to stars, planets, galaxies, groups of galaxies and much more.  The domain of life also offers numerous examples of radical novelty:  from different types of cellular organisms to various species of animals and plants, to humans, society, cultures, and civilizations.  Each of these creations represents a new level of organization that is more powerful than the one that has preceded it and from which it has emerged.  Therefore, all these creations are products of the general process of creation that operates in the universe.

Many scientists have aversion to using the word "creation" because of its religious connotations.  Few researchers use this term as an operational concept.  For this reason, some explanation and demystification of creation is in order to help overcome this aversion.

One can define the process of creation through an operation that produces something new—something that has not existed prior to its emergence.  The distinct feature of creation in this sense is its irreducibility.[98]  We cannot reduce and establish linear causal connection between the level of organization that has emerged and the level of organization from which it has emerged.  We cannot reduce consciousness to biological processes that, according to the theory of the evolution, gave rise to it. We cannot explain how the biological level of organization that sustains insects gives rise to what appears to be intelligent behavior of insect swarms or colonies.[99]

The process of creation has its roots in the very nature of our universe.  Our universe is unique.  It is all there is.  Nothing can come into it from outside and nothing can disappear from it because there is nowhere to disappear.  Therefore, everything must be conserved.  Conservation is the fundamental condition of the existence of the universe.  Without conservation the universe simply cannot exist.

Conservation manifests itself at all levels of organization of reality—from physical to chemical, to biological, and to human social and cultural levels.  Examples of conservation abound.  Exploding a stick of dynamite converts chemical energy into kinetic one.  The transformation of the energy of falling water into electricity is another good example of conservation in action.  Conservation is ubiquitous.  Due to the ubiquity of conservation and its connection to the universe we can view conservation as a truly universal property.

Conservation requires resources.  In the universe of finite objects resources are always limited.  Therefore, access to new resources is the only path to conservation.  Gaining access to new resources requires new possibilities that can only come with new and more powerful levels of organization.  When particles combine into atoms or when atoms form molecules, they acquire a broader range of possibilities, or degrees of freedom.  These new possibilities offer access to new resources that are essential for conservation.  Simple molecules can combine into complex molecules (for example, organic molecules), which expands their possibilities.  Organic molecules allow even more diverse combinations that create new and even more powerful levels of organization.  The emergence of new and increasingly more powerful levels of organization is the very essence of evolution.  Thus, conservation leads to creation and creation leads to evolution.  This process sustains our universe and everything in it.

All systems that exist in the universe consist of parts, or subsystems.  Interactions among subsystems sustain systems and make their existence possible.  These interactions cannot be chaotic; they and the system as a whole require stability.  Therefore, all systems require regulation that ensures stability and a global regulatory mechanism that provides regulation.  The global regulatory mechanism performs a very important function—one that sustains the entire system and all its components.  However, in order to perform this function, the regulatory mechanism must also be stabilized and conserved.

Function is a form of action.  The only way to conserve action is by acting it out.  The more often a function is enacted, the more stable it is and the better it is conserved.  In order to bring a function into an active state, it must be triggered.  This triggering action comes in the form of a signal that can come from the environment in which the function operates; it comes from its connections with other functions or systems.  Therefore, establishing connections with entities in its environment is very important for conserving the regulatory function and the mechanism that performs it.

There are two ways of conserving the global regulatory function:  endogenous and exogenous.  The global regulatory mechanism represents a more powerful level of organization that offers many more possibilities than the level of organization of each subsystem or their sum total. Conservation of a system requires the integration, or equilibration, of the two levels of organization—local and global. In order to come into equilibrium, the global level and the local level of organization need to communicate with each other; in other words, they should have access to each other. Since the global level is more powerful than the local level, it has access to the latter. The local level, however, does not have access to the global level, which presents an obstacle to their equilibration. The overcoming of this obstacle requires the establishment of a frame in which both the global level and the local level of organization are but two particular cases. The integration of the two levels of organization enriches local operations and increases their power. Donald Campbell points to this phenomenon in his discussion of the so-called "downward causality" that involves the influence of the more powerful global level of regulation on operations at the local level.[100] The equilibration of the two levels also leads to the differentiation of the global level. Since the frame that makes this equilibration possible includes both the global and the local level of organization, it represents a new level of organization that is more powerful than both of them. The creation of such frame transcends the capabilities of the system and increases its overall power. With its emergence, the system enters a new cycle in its evolution. Thus conservation requires combination of differences that leads to the creation of a new level of organization and thus the evolution of the system as a whole.

The exogenous way of conserving the regulatory mechanism involves its external connections with other systems and their regulatory mechanism. As has been pointed out, the global function of regulation represents a level of organization that is more powerful than that of each individual subsystem or their sum total. In other words, this function transcends the boundaries of the system and can establish connections with other systems that exist in its external environment. Such connections with external systems also trigger the regulatory function into action, thus helping to conserve it. Eventually, the connections among systems become permanent. The combinations they create result in the emergence of new levels of organization in which systems begin to function as subsystems. These new levels of organization are more powerful than that of each system involved or their sum total. Thus, both the endogenous and the exogenous way of conservation leads to the creation of new and increasingly more powerful levels of organization and, thus, to evolution.[101]

Conservation of function is essential for evolution. Many biologists recognize this fact and use the functional approach in their OOL models. Karl Niklas, for example, maintains that functional traits are objects of conservation.[102] As has been pointed out earlier, the problem with the functional approach used by biologists is that they focus on functions ("privileged functions") that come from the biological realm. Using a biological function to explain the emergence of life from inorganic matter presents an obvious problem. The level of organization that biological functions represent is much more powerful than the levels of organization in the inorganic (physical or chemical) domains. Biological functions simply cannot exist at levels of organization of non-living matter. Using biology to explain processes in the non-organic domain is a form of organicism, or the reduction of the inorganic world to biology. It simply cannot work since the level of organization of physical or non-organic chemical processes is less powerful than the level represented by biological processes. The function that gives rise to life from non-life cannot be limited to biology. It must be related equally to both domains. It must represent a level of organization in which life and non-life are but two particular cases.

Conservation is the only function that is fundamental for both life and non-life. As has been argued earlier, conservation is impossible without creation. This universal process of creation is the common denominator that covers the two such different domains as life and non-life. Therefore, the universal process of creation makes for a good foundation for a comprehensive theoretical perspective on the origin and evolution of early life.

Using the universal process of creation in this capacity is not a subjective or arbitrary decision. This decision passes the test of rational justification and empirical verification. Without creating mental structures in their brain, humans would not be able to experience and interpret reality. Reality for them would be nothing, rather than something. In other words, an act creating these mental structures is a necessary act without which reality would not exist for humans. On the empirical level, we see results of numerous acts of creation all around us. We also have the capacity to create and we have created many new and increasingly more powerful levels of organization that have given rise to many radically new forms in science, art, or society. Therefore, one would be fully justified to use the process of creation as the foundation and the organizing principle of the new perspective on the origin and evolution of early life.

*The Universal Process of Creation and the Evolution of Early Life*

The purpose of this study is not to provide an exhaustive description of the origin and evolution of early life. The completion of this task requires more theoretical and empirical work. The intention of this study is merely to offer a general model for this important transition.

There is no reason to suppose that the origin and evolution of early life is an isolated phenomenon that has nothing to do with the evolution that preceded or followed it. The very notion of evolution presupposes a continuous process where each preceding stage gives rise to the one that follows it. The evolution is a process that has roots in the very nature of the universe. Its fundamental pattern of conservation $\rightarrow$ creation $\rightarrow$ evolution is observable at many levels of organization of reality both before and after the origin of life. The connection between conservation, creation, and evolution has been the subject of several prior studies that have dealt with the cosmic evolution, the biological evolution, and the evolution of culture and civilization.[103] There is no justifiable reason to believe that the universal evolution made an exception in the case of the origin and evolution of early life and abandoned its fundamental pattern only to return to it in the evolution that followed the early life period.

Conservation was the main reason for the process that gave rise to life and then drove the evolution from life's early forms to animals and plants, humans, culture, and civilization.

It is one unified process that connects biological evolution with the evolution of the universe as a whole. Ursula Goudenough points to this singular role of conservation in the origin and evolution of life.

> The transition from prokaryote to protoeukaryote to the last eukaryotic common ancestor (LECA) entailed conservation, modification, and reconfiguration of preexisting genetic circuits via mutation, horizontal gene transfer (HGT), endosymbiosis, and selection, as detailed in previous articles of this collection.[104]

The pattern of conservation → creation → evolution is in evidence in the evolution of early life, particularly in the evolution from prokaryotes to eukaryotes and from eukaryotes to multicellular life. As has been pointed out, the evolution of modern eukaryotic cells was a product of the cell-cell fusion. The symbiosis of archaea and bacteria created a new level of organization that integrated them both as component parts of a single organism.[105] The adaptation of archaea and bacteria to the new and more powerful system that they created produced the modern eukaryotic cell. The prokaryotic bacterium that was engulfed by the archaeon evolved into mitochondria that became an important component of the modern cell involved in energy production.[106]

Conservation played an important role in the evolution of multicellularity. It drove unicellular organisms to aggregate into colonies that preceded the emergence of multicellular organisms. Conservation of functional activities through mutual stimulation was the principal reason for cells to come together. Interactions with each other stimulated individual unicellular organisms, activated their vital functions, and thus contributed to their conservation. These interactions created a new level of organization. Cell-to-cell adherence was the *conditio sine qua non* of multicellularity across all clades; cell-to-cell communication was the basis of multicellular development.[107] Adaptation of individual cells to the new structure led to their functional differentiation, specialization, and cooperation that created multicellular organisms.[108]

Although the emergence of organic molecules capable of replication and the rise of prokaryotic cells needs more theoretical insights and empirical facts, what is known about this process also points to the important role of conservation. The verdict is still out as to whether the emergence of prokaryotic cells was a result of assembly of individual components or, as Kamimura and Kaneko theorize, it was a product of interactions among molecules in clusters that gradually increased their complexity and improved copy fidelity.[109] The scenario proposed by Kamimura and Kaneko appears to be more promising as it solves several problems at the same time: the origin of heredity, metabolism, and compartmentalization.[110] However, no matter what the final verdict is going to be, both paths point to the importance of conservation, as well as adaptation and assimilation as two vital aspects of combining and equilibrating required for creating prokaryotes.

The accumulated knowledge shows that the origin and evolution of early life has many features in common with the general process of creation that operates in the universe. This fact should not come as a surprise since life is a result of the evolution that preceded its emergence. There can be no question that life inherited the properties of the evolution that preceded it and made its emergence possible. All essential features of the general process of creation that propels the evolution of the universe are also observable in all stages and transitions involved in the origin and evolution of early life. Since life has emerged in the course of the evolution of the universe, the fact that it inherited fundamental features of the universal evolution should not be a surprise. Evolution that is worthy of its name can only be continuous and continuity involves heritability. Therefore, there is every reason to view the origin and evolution of early life as an integral part of the universal process of creation.

Just as in the process of the evolution of the universe, assimilation, adaptation, and regulation also play a very important role in the evolution of life. The rise of multicellularity is a good case in point. Many researchers note, for example, that regulation of interactions among cells played a critical part in the emergence of multicellularity. According to Ruiz-Trillo and his team, increased fine-tuning of gene regulation is a key to the successful transition to multicellularity.[111]

Adaptation is another feature that played a major role in the early evolution of life. For example, adaptation of cells to multicellular structures and differentiation of their functions was a very important aspect in this transition. Evolutionary biologists Guy Cooper and Stuart West at the University of Oxford in the United Kingdom, point out in their piece that appeared in *Nature Biology & Evolution* that the process of differentiation of functions as a form of adaptation to new multicellular conditions "was not a consequence but a driver" in the evolution of multicellular organisms.[112]

All current perspectives recognize that the origin and evolution of early life was a continuous process that involved the emergence of radical novelty; that is to say, they recognize that both continuity and discontinuity are involved in the evolution. Yet, they do not show how the two are interrelated and, as a result, cannot resolve the problem of the relationship between continuity and discontinuity in the origin of life.

Instead of resolving this contradiction, they use a palliative approach that assigns priority to either to continuity or discontinuity. In order to explain the emergence of radical novelty, many current OOL perspectives invoke chance or coincidence. The use of chance in explaining radical novelty undermines the very notion that recognizes evolution as an autonomous process that does not require external causes.

By contrast, the new perspective maintains that continuity and discontinuity as two equal and complementary aspects of the evolution. Each new stage in this process has emerged from the one that preceded it as a result of interactions among systems that it sustains. These interactions and the combination of differences that they effect create a new and more powerful level of organization that did not exist prior to its emergence and thus represents discontinuity. Thus the new perspective shows that each new stage emerges from the one that preceded it and, at the same time, it represents a radical novelty that did not exist prior to its emergence. Each new stage conserves and enhances the power of the stage that preceded it. In other words, it shows that continuity and discontinuity are two equal aspects of the same process that complement each other and play an equally important role in the evolution.

Due to their failure to reconcile continuity and discontinuity, the current OOL perspectives cannot create an approach that would be equally applicable to both domains. As a result, they have to assign priority to either life or to non-life. Those who assign priority to life, use either functions ("privileged functions") or forms (common ancestors) that belong to the domain of life. They use the characteristic features of living organisms to explain the abiogenesis in the domain of non-life. In other words, they use results to explain causes. This strategy inverts the cause-effect relationship and uses effects to explain causes, which is not a methodology acceptable in modern science. Those who assign priority to non-life—either self-organization or entropy production models—essentially go to the other extreme. They reduce life to non-life. Prioritizing one realm over another does not and cannot solve the problem. The solution requires a perspective that views life and non-life as two particular cases of a more general common frame.

As a result of the failure to recognize the role of local integrative interactions in creating new and more powerful levels of organization, most current OOL perspectives overemphasize the role of differentiation; they see the evolution primarily as the process in which new organisms branched off from some common ancestor. In other words, they prioritize differentiation over integration. The most vivid example of this prioritization is elaborate phylogenetic trees that show connections of various organisms to their common ancestors.

The available evidence about the process of evolution does not support interpretations that privilege differentiation or, for that matter, integration. The evolution provides no support for prioritizing either one or the other. A good illustration comes, for example, from linguistics. The Indo-European theory of the origin of languages that emphasizes differentiation has not been successful in modeling the actual processes that led to the emergence of new languages. For example, when archeologist and ethnologists tried to identify an ethnic group that spoke the hypothetical proto-Slavic language, their efforts completely failed. Archeological excavations at the location where the theory predicted that the group had resided found no evidence that humans had ever lived in this marshy and inhospitable territory. The available evidence has shown, however, that contrary to the Indo-European theory focused on differentiation, the emergence of major Slavic languages and the ethnic groups that speak them was a result of processes that involved integration and differentiation at the same time: the integration of local tribes led to the differentiation of the major Slavic ethnicities.[113]

Searching for common ancestors and constructing elaborate phylogenetic trees, as many current perspectives do, is a formidable task that produces mixed results. Empirical evidence for the existence of such progenitors is often lacking and may never be available. There is little guarantee that these postulated ancestors have ever existed or that this existence can be proven. As Maureen O'Malley has pointed out, the idea of a common ancestor is probably more about "an ancestral state, a genomic state that we don't know was one single cell."[114] The search for hypothetical progenitors and the construction of elaborate phylogenetic trees may very well be, as some researchers point out, a futile and utterly unhelpful enterprise.[115] The evolution is not about changes from some relatively simple ur-forms, or common ancestors, to more complex forms. The evolution is about transitions from complexity of one kind to complexity of another kind; it is about differentiation and integration working in harmony with each other. The new perspective on the origin and evolution of early life does not prioritize either differentiation or integration. It views differentiation and integration as two equal and complementary aspects of the same process.

The evolution is a universal process. As a universal process, it must have some universal criterion that marks its progress. This criterion must be common to all domains. As has been argued, evolutionary processes in all domains are about creating new and increasingly more powerful levels of organization. The result of the evolution is cascading levels of organization that are nested matryoshka-style in each other. Each new level of organization conserves and enhances the possibilities of the level that precedes it and from which it has emerged. Thus the level of organization is the only criterion that is relevant to all evolutionary processes regardless of a particular domain in which they are taking place.

Levels of organization have received relatively little attention in the studies of the evolution. Although many scientists use this term and can and do identify specific levels of organization, such as subatomic and atomic, molecular, cellular, tissue, organ, organismal, group, population, community, ecosystem, landscape, biosphere, etc., there is no widely accepted definition of what a level of organization is. In the absence of a general and universally accepted definition, this study suggests the following working definition: A level of organization is a network of structurally connected components that has its own distinct level of combinatorial power—that is, it offers distinct degrees of freedom. The characteristic feature of each level of organization is the array of possibilities that it offers. For example, the nervous system in an organism represents a level of organization that has sufficient combinatorial power to regulate biological functions of the organism. Human brain represents another level of organization that is capable of regulating neural functions and through them physiological functions. Brain and mind represent the most powerful level of organization that exists in nature. It derives its power from the capacity to create an infinite number of new and increasingly more powerful levels of organization that are capable of sustaining an equally infinite array of forms and behaviors.

Biologists recognize the existence of levels of organization and even use the term. References to levels of organization are prominent in life sciences; they appear in introductory textbooks, as well as in cutting-edge research. However, the dominant evolutionary theories do not see the level of organization as the principal object of conservation that can be used in explaining the biological evolution. They continue to focus on the more familiar objects of conservation, such as genes, organisms, or some other specific forms.  Moreover, there is also a strong tendency to extend this exclusive approach to other domains—for example, to the cultural or cognitive evolution by using objects of conservation borrowed from biological evolution, such as genes.

There are scientists who point to the universal utility of the level of organization.  Donald Campbell, for example, emphasizes that levels of organization are much more useful in discussing biological evolution than arbitrary schemes of classification.[116]  Etienne Roux and his co-editors see biogenesis as "the progressive constitution of spatial and temporal systems beyond individual proto-metabolic organizations."[117]  Alvaro Moreno approaches prebiotic evolution as the emergence of autocatalytic reaction loops between molecules leading to compartmentalized self-maintained and self-reproducing networks."[118]  Jan Baedke is yet another representative of this trend.  He argues that the perspective that focuses on levels of organization and their origin helps understand better the process of the emergence of such innovations as, for example, multicellularity. [119]

Viewed from the perspective that focuses on specific forms—one that is used in the Darwinian model—evolution appears to be an erratic, unsteady, and ambiguous process punctuated by sudden leaps and reversals that are hard to explain.  By contrast, when viewed from the perspective that focuses on levels of organization, the evolution emerges as a steady and consistent progress.  It shows no exceptions to conservation.  No level of organization has ever become extinct; all levels are conserved. There are no reversals and no instances of non-conservation.  When one views the evolution from the perspective of forms, one can observe that about 95 per cent of all species that have ever lived have become extinct. Viewed from the perspective of levels of organization, one can see that levels of organization that once dominated life world may no longer be dominant; yet they continue to exist.   They have not disappeared but have simply yielded their dominant position to other more powerful levels of organization.  Species that represent these extant levels survive to this day.

The current OOL perspectives lead into another trap that one can describe as chicken-and-egg problem.  For example, in explaining the origin of self-replication, researchers point out that replication requires nucleic acids RNA and DNA for storing information essential in making proteins for replication.  Yet proteins are also necessary for making nucleic acids.  In other words, nucleic acids require proteins, but only nucleic acids can make these proteins.  Both proteins and acids are long molecules that consist of smaller components that are synthesized also by proteins and acids.  Thus, as one researcher aptly characterized the conundrum, contemporary OOL theories end up with what seems to be an irresolvable problem of "two chickens and two eggs."[120]

The proposed new model avoids this trap.  It emphasizes that the proper result of the evolution is not this or that particular form (protein, acid, or organism).  The evolution is not about forms; it is primarily about the emergence of new and increasingly more powerful levels of organization that give rise to new forms.  In other words, the level of organization (one can call it the level of "chickeneggness" for the lack of a better term) that gives rise simultaneously to both the chicken and the egg, or rather many different kinds of chicken and many different kinds of eggs.

As has been pointed out, more powerful levels of organization—that is, levels that regulate interactions among subsystems—conserve all levels that they regulate. They include the essential features of the levels on which they supervene and, thus, conserve them. Humans now dominate Earth. Many species that had at one time been dominant now exist in some extant forms. However, the level of organization that sustains the human race retains all the essential features of all the levels of organization that have preceded it. Nature is frugal and does not like conspicuous consumption. It conserves only what needs to be conserved. What has already been conserved does not need any additional conservation.

This is not to argue against the role of adaptation and selection in the evolution. Indeed, they play an important role but in the processes that take place within levels of organization, not in the transition from one level of organization to another. The capacity of adaptation is primarily a result of the level of organization that offers the possibilities used for adapting.

With this qualification, one can agree with Bonner that natural selection is not a particularly useful concept in explaining the evolution.[121] Each new level of organization that emerges in the course of the evolution conserves everything that has preceded it. Even if some species disappear, the levels of organization that they represented do not. They are conserved. Dinosaurs disappeared long time ago, but their level of organization is alive and well. Species that represent this level are still around and live right next to us.

Finally, when viewed from the perspective based on levels of organization, one can see that the process of the evolution represents a steady increase in power of new emerging levels of organization all the way to human brain that represents the most powerful level of organization in nature. Even though many specific organisms have become extinct, the levels of organization they represented still survive. The evolution of such key aspects of organism as neural organization and modes of behavior reveals steady increase in capabilities and power. Pierre Teilhard de Chardin, a famous French philosopher and paleontologist, was one of the first who pointed to this steady increase in his famous book *The Phenomenon of Man*.[122] The constant growth of the power of the neural system in the course of the evolution is a well-documented fact.[123]

**Conclusion**

The evolution is integral to the world in which we live. Our universe constantly evolves. This evolution makes the existence of the universe and the existence of all it contains possible. The understanding of the evolution is the most important quest in our pursuit of knowledge, both about us and about our place in the universe.

The study of the origin and evolution of early life is perhaps the most intriguing part of this quest. It deals with two domains—the domain of life and the domain of non-life.

There are hardly any two other domains that are more different than these two. We are so sure in our intuitive knowledge of the existence of the fundamental boundary that separates life from non-life that we accept it as self-evident. For us, this difference requires no proof; it just is.

The study of the origin of life defies this intuitive knowledge. It requires a rational explanation of how non-life could give rise to life. It forces us to go against our intuition about the difference between life and non-life and search for their connection. T hat is why the study of the origin and evolution of early life acquires such particular significance. The uncompromising liminal location of the origin of life—between the animate and inanimate world—infringes on established disciplinary boundaries. It requires the recognition of the autonomy of each domain and defies any prioritization of one over the other. It permits no reductionism. Rather than look for similarities between life and non-life, the new theoretical perspective must focus and combine their differences without reducing them.

Most current perspectives on the origin and evolution of early life accept the Darwinian model of the evolution. They rest on the foundation of Darwin's theory. Yet the origin and evolution of early life in many ways challenge this connection. The core of the Darwinian model is variability and natural selection. According to this model, natural selection acts on variations that are due to imperfections in replication or to chance. The more adaptable variations survive and bring about evolutionary advances. Natural selection rewards the capacity of an organism to adapt to new conditions.

This model does not really work well in the domain of non-life. The capacity for adaptation is a distinct feature of life and biological organisms. We cannot really talk about adaptability of inanimate matter. Atoms certainly do not change when they combine into molecules; they remain atoms regardless of whether we find them in the Sun, in interstellar space, or here on Earth.

If there is no variability, natural selection has nothing to act upon—it simply does not work; and if there is no natural selection, the Darwinian model is not applicable to the inanimate world. How then can a radical novelty, such as life, emerge from this world that has no variation on which natural selection can work? The Darwinian model is completely silent on this matter and simply resorts to various miraculous devices such as chance or coincidence. That is perhaps the reason why Bonner is skeptical about the universality of natural selection. He recognizes its applicability to later stages in the evolution of animals and plants, but not to the evolution of early life. He writes:

> In unicellular microorganisms, which at one time in early Earth history were probably the only living eukaryotic forms, *natural selection plays a relatively minor role*, but with size increase, first made possible by the invention of multicellularity, selection plays an increasingly central role in evolutionary change.[124]

In a way, Bonner alerts us to a possibility that the Darwinian model may not have a universal application and a broader view is necessary to explain the general evolutionary process that led to the emergence of life.

The origin and evolution of early life requires a broader frame than the Darwinian model that originated in the study of life and deals mostly with the biological realm. Attempts to apply it to other domains—social, cultural, intellectual, etc.—have produced mixed result.[125] Its applications to the evolution of the universe and the inorganic world in general are practically non-existent.

Using the Darwinian model in fields other than biology poses a difficult choice: either one has to reduce a non-biological evolution to biological processes, as the Darwinian model often does (for example, reducing evolution of human behavior, society or culture to genetics), or one has to claim that each individual domain has its own unique evolutionary process that has little to do with evolutionary processes operating other domains. One route leads to reductionism; the other leads to the proliferation of dramatically different evolutionary models. This latter route undermines the unity of the evolution since evolution makes sense only as a unitary process. If we are to take the evolution seriously, we must recognize that despite some specific differences, the evolution can exist only as a continuous and unitary process. The evolution of life does not start from point zero; it has its roots in the evolution of the inanimate world that precedes it. If we break this vital connection, we have to reject the very notion of evolution.

The new perspective presented in this study does not prioritize the domain of life over the domain of non-life. This perspective rests on the recognition of the universal process of creation that propels the evolution of the universe and everything in it. The choice of the process of creation as the main organizing principle of the new perspective is not subjective or arbitrary. It passes the test of rational justification and empirical verification. Without the process of creation humans would not be able to make any sense of reality. Only mental constructs that humans create in their mind make their perceptions of reality and interpretations of these perceptions possible. The universe is full of examples of wonderful creations that represent increasingly more powerful levels of organization. Humans inherited this capacity to create in the course of the evolution that has led to the emergence of humanity. Our entire civilization is the most vivid example of the power of the process of creation.

The universal process of creation covers both the domain of life and the domain of non-life. This process is a natural result of conservation that is ubiquitous in the universe and is the main condition of its existence. Conservation requires creation and creation results in evolution. Indeed, the exhaustive description of the transition from non-life to life requires more theoretical and empirical research that will fill remaining lacunae and provide more evidence. However, the sequence of conservation → creation → evolution is as true for the inanimate world as it is true for the animate one. It is a bridge between the domain of life and the domain of non-life.

Since the process of creation is broad enough to be relevant to the entire universe and everything in it, the model based on this process also has a broad applicability. Evolutions in all domains follow the same pattern of conservation → creation →evolution. For this reason, the model of the evolution based on the process of creation is applicable in domains other than biology. In addition to biology, it can be used in interpretations of the evolution of cosmos, culture, society, intellectual evolution, and others.[126]

Both life and non-life conserve themselves by creating new and increasingly more powerful levels of organization. This is not to deny the difference between life and non-life. As many biologists have pointed out, the critical difference between life and non-life is autonomy. Ruiz-Murazo and his co-authors, for example, see autonomy as the distinct feature of life.[127] New and increasingly more powerful levels of organization emerge in the inanimate world and, obviously, inanimate matter is involved in these creations. However, inanimate matter is subject to external forces, such as gravity, magnetism, etc. The drive for conservation in living organisms is largely internal. Just like non-life, life continues the process of creating new and increasingly more powerful levels of organization. However, living organisms have their own internal capacity for securing their own resources for performing this creative process and thus conserving themselves. They are the source of their drive that starts with conservation and leads to creation and evolution.

The new perspective offers a possibility to resolve the problem of the origin of life without resorting to chance or coincidence. The suggested general path does not require any miracles. Aggregations of organic molecules with topological differentiation into the inner and outer regions, their adaptation to this structure, differentiation and specialization of their functional activities could be the path to the origin of the level of organization that gives rise to autonomous entities capable of storing information, as well as harvesting and transmitting available ambient energy from the environment to their inner zone.

This scenario does not require any chances or coincidences. It describes the same processes of aggregation, differentiation, and specialization that we can see in the inorganic world. The only difference is that there is no need for external forces of compulsion. They occur as a result of the internal momentum to conserve.

Another advantage of the new perspective is that it eliminates contradictions associated with binary opposites, such as continuity vs. discontinuity, differentiation vs. integration, determinism vs. indeterminism. The process of creation reconciles these binary opposites. It shows that they are not ontological opposites but are equal and complementary aspects of the same process of creation. The connection between these aspects plays an essential role in the process of creation and hence in the evolution.

The universal evolution is a unitary process. It conserves one important function that sustains the universe and everything in it: the capacity to create new and increasingly more powerful levels of organization.  All stages of the universal evolution conserve this important function by exercising it in different variations over and over again.  Humans have inherited this function in the course of the evolution that is the source of their existence.  The capacity to create new and increasingly more powerful levels of organization represents the fundamental promise of their future existence.  By performing this function, humans fulfill their destiny—the destiny that is eternally etched in the evolution that made the existence of the human race possible.

**NOTES**

Erin Garcia de Jesus, "Two New Books Investigate Why It's so Hard to Define Life." *Science News*, March 17, 2021, https://www.sciencenews.org/article/biology-two-new-books-why-hard-define-life-meaning.

Nick Lane, *The Vital Question:  Energy, Evolution, and the Origins of Complex Life* (New York:  W. W. Norton & Company, 2015), p. 6, http://archive.org/details/vitalquestionene0000lane.

E. Camprubí, J. W. de Leeuw, C. H. House, F. Raulin, M. J. Russell, A. Spang, M. R. Tirumalai, and F. Westall, "The Emergence of Life," *Space Science Reviews*, vol. 215, no. 8 (December 12, 2019), p. 34, https://doi.org/10.1007/s11214-019-0624-8.

Kepa Ruiz-Mirazo, Juli Pereto, and Alvaro Moreno, "A universal definition of life: Autonomy and open-ended evolution," *Origins of Life and Evolution of Biospheres*, vol. 34 (2004) 34, pp. 323–46; p. 323.

William Bains, "Getting Beyond the Toy Domain:  Meditations On David Deamer's "Assembling Life," *Life,* vol. 10, no. 18 (2020), p. 1.

Carl Sagan, "Life," *The Encyclopedia Britannica* (London:  William Benton, 1970); Ruiz-Mirazo et al., "A universal definition of life," p. 327.

NASA, "About Life Detection," *Astrobiology at NASA/Life in the Universe*, https://astrobiology.nasa.gov/research/life-detection/about.

Claus Emmeche, "Defining Life as a Semiotic Phenomenon," *Cybernetics and Human Knowing*, vol. 5 (1998), pp. 3–17;  Ruiz-Mirazo et al, "A Universal Definition of Life," p. 326.

R. Shapiro and G. Feinberg, "Possible Forms of Life in Environments Very Different from the Earth," in J. Leslie (ed.), *Physical Cosmology and Philosophy* (New York:  MacMillan, 1990), pp. 248-55.

F. Varela, F. J.: 1994, "On Defining Life," in G. R. Fleischaker, S. Colonna and P. L. Luisi (eds.), *Self-Production of Supramolecular Structures: From Synthetic Structures to Models of Minimal Living Systems* (Dordrecht:  Springer, 1994), pp. 23–33; F. Varela and H. Maturana, *Autopoiesis and Cognition:  The Realization of the Living* (Dordrecht:  D. Reidel Publishing Company, 1972); F. Varela, H. Maturana, R. Uribe, "Autopoiesis:  the organization of living systems, its characterization and a model," *Currents in Modern Biology*, vol. 5 (1976), pp. 187-96, PMID 4407425 DOI: 10.1016/0303-2647(74)90031-8.

Ruiz-Mirazo, "A universal definition of life," p. 330.

Addy Pross, "Physical Organic Chemistry and the Origin of Life Problem:  A Personal Perspective," *Israel Journal of Chemistry*, vol. 56, no. 1 (January 2016), pp. 83–88, https://doi.org/10.1002/ijch.201500073, p. 87.